\documentclass[shortAfour, sagev, doublespace, times]{sagej}

\usepackage{moreverb,url}

\usepackage[colorlinks,bookmarksopen,bookmarksnumbered,citecolor=red,urlcolor=red]{hyperref}
\usepackage{graphicx, natbib} % Allows including images
\usepackage{booktabs} % Allows the use of \toprule, \midrule and \bottomrule in tables

\usepackage{amsfonts,amssymb, amsmath,breqn, mathrsfs}
\usepackage[ruled, vlined]{algorithm2e}
\usepackage{threeparttable}
\usepackage{caption}
\usepackage{threeparttable}
\usepackage{array}
\usepackage{tabularx}

\newcommand{\boldX}{\mathbf{X}}
\newcommand{\boldH}{\mathbf{H}}
\newcommand{\boldXbar}{\overline{\boldX}}
\newcommand{\boldWbar}{\overline{\boldW}}
\newcommand{\boldW}{\mathbf{W}}
\newcommand{\boldw}{\mathbf{w}}
\newcommand{\boldF}{\mathbf{F}}
\newcommand{\boldS}{\mathbf{S}}
\newcommand{\boldphi}{\boldsymbol{\Phi}}

\newcommand{\Fstar}{F^*}
\newcommand{\Sstar}{S^*}

\newcommand{\Rstar}{R^*}
\newcommand{\gopt}{g^{opt}}

\newcommand{\summationT}{\sum_{a_T=1}^{K_T}}
\newcommand{\summationj}{\sum_{a_j=1}^{K_j}}

\newcommand{\muaipwhat}{\hat{\mu}_{j,a_j}^{AIPW}}
\newcommand{\pij}{\pi_{j,a_j}}

\newcommand\BibTeX{{\rmfamily B\kern-.05em \textsc{i\kern-.025em b}\kern-.08em
T\kern-.1667em\lower.7ex\hbox{E}\kern-.125emX}}

\begin{document}

\runninghead{Zhong and Wang}

\title{Augmented Patient Preference Incorporated Reinforcement Learning (APP-RL) to Estimate the Optimal Dynamic Treatment Regime}

\author{Yingchao Zhong\affilnum{1} and Lu Wang\affilnum{1}}

\affiliation{\affilnum{1}Department of Biostatistics, University of Michigan, Ann Arbor, Michigan, USA}

\corrauth{Yingchao Zhong, Current affiliation: Division of Cardiology, California Pacific Medical Center, Sutter Health, San Francisco, California 94109, USA. }

\email{zhongych@umich.edu}

\begin{abstract}
Dynamic treatment regimes (DTRs) are sequential decision rules that individualize treatments to each patient at each treatment stage adapting to their past clinical course. Existing literature typically accommodates each individual’s medical history, but overlooks a patient’s preferences. We propose a method that incorporates a patient's latent preferences through data augmentation into a tree-based reinforcement learning method to estimate optimal dynamic treatment regimes for multi-stage, multi-treatment settings. For each patient at each stage, we derive the posterior distribution of preferences given responses to a questionnaire, and then subsequently weight multiple outcomes with the estimated preferences to identify the optimal stage-wise personalized decision. For multiple stage situations, we grow a decision tree at each stage and implement the algorithm recursively using backward induction. Our proposed method, named Augmented Patient Preference incorporated Reinforcement Learning (APP-RL) is robust, efficient, and leads to interpretable DTR estimation. The finite-sample performances of the proposed method has been thoroughly evaluated through simulation studies. 
\end{abstract}


\maketitle

\section{Introduction} \label{s:introduction}

Personalized health care aims to predict patients' responses to targeted therapy using patient characteristics through a multifaceted approach \citep{hamburg2010path}. Instead of one-size-fits-all, personalized medicine hopes to concentrate therapeutic interventions on ``those who will benefit, sparing expense and side effects for those who will not'' \citep{national2011toward}. Patient responses to treatments may vary due to different levels of heterogenities, such as genetics, environmental factors, and the interplay between the two. Because of this, an appropriately personalized treatment plan needs to be sensitive and adaptive to a patient's evolving condition, especially in the case of chronic diseases. Dynamic treatment regimes (DTRs) are sequences of treatment decision rules, in which treatment decisions are adapted over time in response to an individual's treatment response and trajectory \citep{chakraborty2013statistical, chakraborty2014dynamic, murphy2003optimal}. With data-driven adaptation of the reinforcement learning problem \citep{sutton2018reinforcement}, DTRs play an important role in evidence based medicine by mathematically formulating the complicated problem of making decisions dynamically at multiple stages in order to maximize a clinically meaningful reward \citep{moodie2012q}. 

A large number of methods have been proposed to evaluate the optimal DTR, including Q- and A-learning \citep{murphy2003optimal, schulte2014q}, marginal structural model with inverse probability weighting (IPW) \citep{robins2000marginal}, G-estimation of structural nested mean models \citep{robins2004optimal}, and other likelihood based methods \citep{thall2007bayesian}. In more recent years, machine learning flavored methods have joined the arsenal of available methods for optimizing DTRs, including tree-based methods \citep{laber2015tree, tao2017adaptive, tao2018tree}, and list-based method \citep{zhang2018interpretable}. Despite the abundant selection of methods to find optimal DTRs, all stated methods rely on pre-specifying a single metric of interest (e.g. survival time, adjusted quality of life, tumor response, etc), thereby forcibly simplifying complex medical scenarios in the formulation of the problem.

Evidence based optimization of one outcome further ignores an important and personalizable part of a patient's experience. In reality, clinical decisions often result in a plethora of outcomes in competing directions of desirability to the patient. The hand surgery field provides an illustrative example. In light of the current opioid epidemic, surgical specialties are re-examining current post-operative opioid prescription habits. Patients have to weigh between the benefits (strong pain relief) and risks (addiction, side effects, etc) of using opioid medications vs non-opioid pain relief (weaker pain relief, but potentially fewer risks). In this case, patient preference could help guide whether to prescribe opioid medications. Similarly, we can illustrate the challenges with another example from neurology. Anti-epileptic drugs (AEDs) often come with side effects such as sedation, somnolence, distractibility, insomnia, and dizziness \citep{ortinski2004cognitive}. Even with all of these side effects, patients might willingly tolerate their AEDs if these drugs control their seizures sufficiently such that they could retain more autonomy, (i.e. by maintaining driving privileges) \citep{krumholz2009driving}. In this case, patients have to balance between value for autonomy (i.e. lower risk of seizures but higher burden of side effects) with their need for mental clarity (higher risk of seizures, but more mental clarity). This example illustrates a complex system of pros and cons that patients need to navigate using their value system, previous experience, and changing needs. A college student in an urban college environment might be willing to forgo his driving privileges in return for mental acuity to meet the demands of school, only to reverse the decision after moving to the suburbs, while a patient with a history of opioid addiction might opt for non-opioid pain medications for fear of relapse. In both of these cases, optimizing one outcome (pain relief or reducing risk of seizure) fits neatly within the evidence based paradigm, but the inclusion of patient preference to coordinate between multiple outcomes does not. Appropriate coordination using patient preference then necessitates the search for another framework.

In recent decades, the medical community has also recognized this gap and has shifted from a paternalistic approach to advocating for patient input through a shared decision-making (SDM) framework \citep{barry2012shared, basu2007value}. Augmentation with patient preferences and values into decision making contributes to a more holistic approach to patient care. A survey of SDM literature shows that patient input has a positive correlation with satisfaction scores and quality of life outcomes \citep{kashaf2015does, shay2015evidence}. SDM has also been shown to reduce costs with unnecessary procedures \citep{oshima2013shared}, thus making it an actionable policy for reducing cost. Despite these documented benefits from the literature, physicians have struggled to practice shared decision making, partly because patient preference is challenging to quantify and incorporate. Current popular shared decision making approaches include using decision aids and increased communication between the clinician and patient \citep{barry2012shared}. Diverse approaches cited here are difficult to model, making it challenging to draw conclusions from data. 

We propose a modeling approach to incorporate patient preference. To properly optimize an outcome representative of patient preference, accurate estimation of the preference itself is paramount. We endorse modeling preference as a latent variable and estimating it through an item response approach \citep{embretson2013item}. Patients could communicate their preference through responses on a questionnaire. These questions may ask patients to rate their agreement with certain statements, or ask patients to rate the importance of certain activities in their lives. Butler et. al \cite{butler2018incorporating} used this approach to estimate patient preference from surveys with binary $\{0,1\}$ responses. Their method, designed to select the optimal treatment between two potential choices at one decision time point, combined estimated preferences along with Q-learning to find the optimal individualized treatment regime (ITR). 

For our scenario, chronic diseases require treatment plans to adapt to patient trajectories over a changing disease course. Furthermore, as the number of treatment stages increases, the number of treatment combination options often also increases, making methods with only two treatment options limiting for realistic settings. As of writing, the authors are unaware of methods in the literature that are able to accommodate multiple stages with multiple treatment options while considering patient preference. In this paper, we propose a method that augments the combination of two potentially competing outcomes with estimated patient preference. As an illustrative example, we will consider competing outcomes of efficacy and toxicity. Our proposed method accommodates the selection between more than two treatments per stage, and allows patient preference to evolve through the stages. Finally, we propose modeling preference through a polytomous latent variable model \citep{moustaki2000generalized, bartholomew2011latent}, which allows us to estimate preferences more precisely through questions with categorical responses, an extension from binary responses in Butler et al \cite{butler2018incorporating}. We combine competing outcomes with a linear utility function weighted by estimated patient preferences, and seek for the treatment decision that would provide the highest patient satisfaction using a tree-based reinforcement learning (T-RL) method. The remainder of this paper is organized as follows. In Section \ref{s:methods}, we introduce the notation, problem and goal, followed up Section \ref{s:optimization}, where we describe our estimation procedure and required assumptions. In Section \ref{s:algorithm}, we describe the implementation details. We evaluate our method in Section \ref{s:simulation} through simulations across multiple scenarios. Finally, we conclude with discussions and ideas for future directions in Section \ref{s:discussion}. 

\section{Patient Preference Incorporated Dynamic Treatment Regimes} \label{s:methods}
Let $T$ denote the number of treatment stages, and let $K_j$ be the number of treatment options at the $j^{th}$ stage. Let $A_{ij}$ denote the observed treatment for the $i^{th}$ patient at the $j^{th}$ stage. Note that when referring to a specific treatment, we will drop the subscript $i$, i.e. treatment $a_j$. Let $\mathbf{X}_{ij}$ denote patient characteristics prior to treatment assignment at stage $j$. In addition, we assume that each patient has an evolving preference $H_{ij}$, which can be derived from answers $\mathbf{W}_{ij}$ to a questionnaire at stage $j$. Finally, we assume that each patient will have two observed outcomes at each stage, efficacy $F_{ij}$ and toxicity $S_{ij}$. In general, we denote all history, or history up to stage $K$ for a given variable with an overhead bar (i.e., $\overline{\mathbf{W}}_{i}$ and $\overline{\mathbf{W}}_{iK}$, respectively).

We will assume a utility of the form $U(F,S; H) = \Phi(H)F+\{1-\Phi(H)\}S$ to designate the utility function at each stage, where $\Phi(\cdot)$ denotes the cumulative distribution function of a normal random variable. The choice of utility function is flexible, but assuming a linear weighted sum utility function is a common approach in multiobjective optimization \citep{marler2004survey, lizotte2012linear}. This utility is also intuitive in that a patient with preference $H$ cares $\Phi(H)/\{1-\Phi(H)\}$ more about $F$ than about $S$. Let the overall outcome of interest that we would like to optimize be $Y=f(U_1,\ldots,U_T)$, where $f(\cdot)$ is a pre-specified function. We assume that $Y$ is bounded, and that higher values are more desirable. Going forward, we will proceed with $f(\cdot)$ as the sum, i.e. the sum: $Y = U_1+\ldots+U_T$, although other functions of $f$ could also be optimized in a similar manner. 

 Let $g_j(\boldXbar_{ij},\boldWbar_{ij})$ be a function that maps from covariate and survey history to the domain of treatment assignment $A_{ij}$. The expected potential reward of stage specific decision rule $g_j(\boldXbar_{ij},\boldWbar_{ij})$ for patient $i$ is therefore defined as $\mathbb{E}\left[\sum_{a_{j}=1}^{K_j} \{\Phi(h_{ij})F_{ij}^*(a_j)+[1-\Phi(h_{ij})]S_{ij}^*(a_j) \} I\{g_j(\boldXbar_{ij},\boldWbar_{ij})=a_j\}\right]$, where $\Fstar_{ij}(a_j) = \Fstar_{ij}(A_{i1},\ldots,A_{i,j-1},a_{j})$ denotes the counterfactual outcome where the patient is assumed to have taken treatment $a_j$ at stage $j$, conditional on previous treatment decisions $A_{i1},\ldots, A_{i,j-1}$, and equivalently for $\Sstar_{ik}(a_j)$. Our goal is to find a sequence of individualized decision rules, $\mathbf{g(\boldXbar_{i},\boldWbar_{i})}=\left[g_1(\boldXbar_{i1},\boldWbar_{i1}),\ldots,g_T(\boldXbar_{iT},\boldWbar_{iT})\right]$, that optimize the potential outcome of $Y_i$. For the sake of brevity going forward, let us abbreviate $g_j(\boldXbar_{ij},\boldWbar_{ij})$ with $g_j$ and drop the patient index $i$ when there is no room for confusion. 

\section{Optimization of $\mathbf{g}$ using iterative Augmented Patient Preference incorporated Reinforcement Learning (APP-RL)} \label{s:optimization}
\subsection{Bridging the counterfactual framework to the observational data}
\textbf{Stage $\mathbf{T}$}

At stage $T$, the counterfactual potential reward under decision rule $g_T$ and conditional on previous treatments $(A_{1}, \ldots, A_{T-1})$ is $R_{T}^*(g_T) = \Phi(H_{T})F_{T}^*(g_T) + \{1-\Phi(H_{T})\}S_{T}^*(g_T)$, where $F_{T}^*(g_T) = \sum_{a_T=1}^{K_T} F_{T}^*(a_T) I(g_T=a_T)$ and $S_{T}^*(g_T)= \sum_{a_T=1}^{K_T} S_{T}^*(a_T) I(g_T=a_T)$. The performance of $g_T$ is measured by the expected counterfactual outcome $E\{R_{T}^{*}(g_T)\}$. The optimal rule, $g_T^{opt}$, then satisfies $E\{R_{T}^{*}(g_T^{opt})\} \geq E\{R_{T}^*(g_T)\} \forall \ g_T \in \mathscr{G}_T$, where $\mathscr{G}_T$ is the class of all potential regimes. 

In order to relate the observed data to counterfactual outcomes \citep{murphy2001marginal, orellana2010dynamic, robins2009estimation}, we make the following assumptions. 
\begin{enumerate}
    \item Consistency: the observed outcome is the same as the counterfactual outcome under a patient's received treatment, i.e., 
    $F_{T} = \sum_{a_T=1}^{K_T} F_{T}^*(a_T) I\{A_{T}=a_T\}$ and $S_{T} = \sum_{a_T=1}^{K_T} S_{T}^*(a_T) I\{A_{T}=a_T\}$ 
    
    \item No unmeasured confounding: treatment $A_{T}$ is randomly assigned with probability possibly dependent on $\boldXbar_{T}$ and $\boldWbar_{T}$, i.e.,
    $\{\Fstar_{T}(1),\ldots,\Fstar_{T}(K_T) \} \perp A_{T} \vert \boldXbar_{T}, \boldWbar_{T}$, and  $\{\Sstar_{T}(1),\ldots,\Sstar_{T}(K_T) \} \perp A_{T} \vert \boldXbar_{T}, \boldWbar_{T}$
    
    \item Positivity: There exists constants $0 < c_0 < c_1$ such that, with probability $1$, the propensity score $\pi_{a_T}(\boldXbar_{T},\boldWbar_{T}) = Pr(A_{T}=a_T \vert \boldXbar_{T},\boldWbar_{T}) \in (c_0, c_1)$
     
     \item Latent variable independence: $H_{T} \perp (A_{T}, F^*_{T}(a_T), S^*_{T}(a_T)) \vert \boldXbar_{T}, \boldWbar_{T}$ 
\end{enumerate}

The first three assumptions are standard assumptions in causal inference used to connect observed data with counterfactual framework. The last assumption is proposed to facilitate separate modeling of preference and outcomes, but can be weakened at the expense of more complicated models and estimation procedure \citep{butler2018incorporating}. 

Notice that we can estimate the left hand side using the following:
 \begin{multline} \label{eqn1}
     \hat{E}\left[R_{T}^*(g_T)\right] = \mathbb{P}_n \Bigg[ \summationT \Big[ E\left\{\Phi(H_{T})\vert \boldXbar_{T}, \boldWbar_{T} \right\} \mu_{T, a_T}^F(\boldXbar_{T}, \boldWbar_{T}) \\
     +E\left\{1-\Phi(H_{T})\vert \boldXbar_{T}, \boldWbar_{T} \right\} \mu_{T, a_T}^S(\boldXbar_{T}, \boldWbar_{T})\Big] I\{g_T=a_T\}  \Bigg],
\end{multline}
where $\mu_{T, a_T}^F(\boldXbar_{T}, \boldWbar_{T}) = E(F_{T} \vert A_{T}=a_T, \boldXbar_{T}, \boldWbar_{T})$, and likewise $\mu_{T, a_T}^S(\boldXbar_{T}, \boldWbar_{T})= E(S_{T} \vert A_{T}=a_T, \boldXbar_{T}, \boldWbar_{T})$. Note that the left hand side (LHS) of Equation (\ref{eqn1}) is expectation of potential outcome, which can be estimated using only observed data as shown by the right hand side (RHS). Furthermore, note that we can separately model preference and outcomes using assumption (4). To find the optimal regime, we want to find $\gopt_T = arg \ max_{g_T \in \mathscr{G_T}} $ RHS of Equation(\ref{eqn1}). 

\textbf{Stage $\mathbf{j}$}

At stage $j$, $T-1 \geq j \geq 1$, $\gopt_j$ can be derived from the observed data via backward induction. Here, $\boldphi(\cdot)$ represents vectorized version of $\Phi(\cdot)$. Following presumption of maximizing the summation of stage-specific rewards from stage $j$ onward, we define the following stage $j$ reward, which is a cumulative sum of stage $j$ to $T$ utility functions:

$R_{j}^{*}(a_j) = \boldphi(\boldH_{j+}) \otimes \boldF_{j+}^{*}(a_j) + \{1-\boldphi(\boldH_{j+})\} \otimes \boldS_{j+}^{*}(a_j)$,
where \\
$\boldphi(\boldH_{j+}) = [\Phi(H_j), \Phi(H_{j+1}), \ldots, \Phi(H_{T})]$, $\boldF_{j+}^{*}(a_j) = [\Fstar_{j}(a_j) \ \Fstar_{j+1}(a_j) \ \ldots \Fstar_{T}(a_j)]$, $\boldS_{j+}^{*}(a_j)= [\Sstar_{j}(a_j) \ \Sstar_{j+1}(a_j) \ \ldots \Sstar_{T}(a_j)]$, and $\otimes$ denotes the dot product. Note that for $k=j$, $\Fstar_{k}(a_j)$ and $\Sstar_{k}(a_j)$ are as defined previously and for $k > j$, $\Fstar_{k}(a_j) = \Fstar_{k}(A_{1},\ldots,A_{j-1},a_j, g_{j+1}^{opt},\ldots, g_{k}^{opt})$ denotes a counterfactual outcome given future optimized treatments and conditional on $A_{1},\ldots, A_{j-1}$ and taking treatment $a_j$ at stage $j$ (similarly for $\Sstar_{k}(a_j)$). Then the optimal regime at stage $j$ satisfies $E\{R_j^{*}(\gopt_j)\} \geq E\{R_j^{*}(g_j)\}$ for all $g_j \in \mathscr{G}_j$, where $\mathscr{G}_j$ is the class of all potential regimes at stage $j$.

As with stage T, we make similar assumptions to link observed data to their counterfactual versions. We make assumptions of consistency, where our counterfactual outcome matches our observed outcome given a specific treatment $a_j$. We assume no unmeasured confounding, where given measured confounders, the choice of treatment $A_j$ is randomly assigned. Similarly, we assume positivity, i.e. $\pi_{a_j}(\boldXbar_{j},\boldWbar_{j}) = Pr(A_{j}=a_j \vert \boldXbar_{j},\boldWbar_{j})$ is bounded away from zero, and latent variable independence: $H_{k} \perp (A_{j},F^*_{k}(a_j),S^*_{k}(a_j)) \vert \boldXbar_{k}, \boldWbar_{k}$, where $k \geq j$.

Combining the above,
\fontsize{10}{12}
\begin{multline} \label{eqn2}
    \hat{E}[R_{j}^*(g_j)] = 
    \mathbb{P}_n \Bigg[ \summationj \Big[ E\left\{\Phi(H_{j})\vert \boldXbar_{j}, \boldWbar_{j} \right\} \mu_{j, a_j}^F(\boldXbar_{j}, \boldWbar_{j})+E\left\{1-\Phi(H_{j})\vert \boldXbar_{j}, \boldWbar_{j} \right\} \mu_{j, a_j}^S(\boldXbar_{j}, \boldWbar_{j})\Big] I\{g_j=a_j\}  \Bigg] \\ 
    + \sum_{t=j+1}^{T} \mathbb{P}_n \left[ \sum_{a_j=1}^{K_j} \Big[ E \{ \Phi(H_{t}) \vert \boldXbar_{t}, \boldWbar_{t} \}  \mu_{t,a_j}^{F}(\boldXbar_{j}, \boldWbar_{j}) + E\{1-\Phi(H_t) \vert \boldXbar_{t}, \boldWbar_{t} \}\mu_{t,a_j}^{S}(\boldXbar_{j}, \boldWbar_{j}) \Big] I\{g_j=a_j\} \right],
\end{multline}
\normalsize

where $\mu_{j, a_j}^F(\boldXbar_{j}, \boldWbar_{j}) = E(F_{j} \vert A_{j}=a_j, \boldXbar_{j}, \boldWbar_{j})$ and $\mu_{t,a_j}^{F}(\boldXbar_{j}, \boldWbar_{j})$ denotes  \\$E\left[F_{t}(A_1,\ldots, A_{j}, g_{j+1}^{opt}, \ldots, g_t^{opt}) \vert A_{j}=a_j, \boldXbar_{j},\boldWbar_{j}\right]$ (equivalently for $\mu_{j, a_j}^S(\boldXbar_{j}, \boldWbar_{j})$ and $\mu_{t,a_j}^{S}(\boldXbar_{j}, \boldWbar_{j})$). Again the RHS can be estimated from observed data only and is a combination of separately estimated preference and outcomes. Under these assumptions, the optimization problem at stage $j$, among all potential regimes $\mathscr{G}_j$, can be written as
$\gopt_j = arg\ max_{g_j \in \mathscr{G}_j}$ RHS of Eqn (\ref{eqn2}). 

\subsection{APP-RL for observational data based optimization}

Our proposed method, named Augmented Patient Preference incorporated Reinforcement Learning (APP-RL), is summarized as follows. Through iteration and sequential estimation, APP-RL is able to combine elements of patient preference and observed outcomes to get the optimal decision rules for each stage. APP-RL first uses survey information from each stage to estimate patient preferences through an expectation maximization algorithm. APP-RL then combines predicted preferences with observed outcomes of toxicity and side effects into previously mentioned utility function and uses a tree-based reinforcement learning method to find the optimal decision rule for that stage. Finally, APP-RL moves backwards through the stages, obtaining optimal decision rule first for stage $T$, next for stage $T-1$, etc, and lastly for stage $1$. 

Although the tree-based reinforcement learning component has some similarities to traditional CART methods, there are important differences that separate the two. Traditional CART methods are supervised learning methods that repeatedly split a parent node into child nodes, generally resulting in purer (fewer misclassification) nodes. Commonly used purity measures include the Gini index, information gain, and least squares deviation \citep{breiman2017classification}. In this framework, each observation carries a label, and the goal of the CART method is to use covariates to correctly classify each subject with its observed label. In contrast, the estimation target of dynamic treatment regime problem, the optimal treatment, is not observed (a patient often does not get the most optimal treatment). Rather, the optimal treatment needs to be inferred indirectly from other patients' treatments and response trajectories. Instead of trying to classify correctly each patient to their assigned treatment, our goal is to optimize the counterfactual mean outcome for the every patient. Aligned with this goal, we propose to use an augmented inverse probability weighted (AIPW) estimator for the counterfactual outcome to be used within our purity measure.

\subsubsection{APP-AIPW estimators}~\\
Our APP-AIPW estimator based off of the AIPW estimator \citep{rotnitzky1998semiparametric}, a doubly robust and consistent estimator which in our case takes into account both observed outcome and estimated patient preference. 

For stage $T$, the APP-AIPW estimator for $E\{ \Rstar_{T}(g_T) \} = \mathbb{P}_n \left[\frac{I(A_{T}=g_T)}{\hat{\pi}_{T,A_T}(\boldXbar_{T},\boldWbar_{T})}R_{T} + \{1- \frac{I(A_{T}=g_T)}{\hat{\pi}_{T,A_T} (\boldXbar_{T},\boldWbar_{T}) }\} \hat{\mu}_{T,g_T}(\boldXbar_{T},\boldWbar_{T}) \right]$, where $R_{T} = \Phi(\tilde{h}_{T})F_{T}+\{1-\Phi(\tilde{h}_{T})\}S_{T}$, where $\tilde{h}_T$ is the estimated preference. $\hat{\mu}_{T,g_T}(\boldXbar_{T},\boldWbar_{T})$ can be any model of $R_{T}$ as a function of observed covariates and survey data accumulated by stage $T$. Under previously listed causal assumptions, if either the propensity model $\pi_{T,A_T}(\boldXbar_{T},\boldWbar_{T})$ or the conditional model $\mu_{T,g_T}(\boldXbar_{T},\boldWbar_{T})$ are correctly specified, then this estimator is a consistent estimator for $\hat{E}\{ \Rstar_T(g_T) \}$.

For stage $j < T$, the stage-specific APP-pseudo-outcome analog of the stage $j$ reward is $\widetilde{PO}_{j} = R_{j}(A_1, \ldots, A_{j-1}, A_j, g_{j+1}^{opt}, \ldots, g_T^{opt})$, which is a weighted combination of stage-specific pseudo-outcomes summed across stages $j$ to $T$. In this formulation, we add the observed value of stage $j$ (including an estimated value of $h_{j}$), and follow the convention in Huang et al\cite{huang2015optimization}, where instead of only using the model-based values under optimal future treatments, we use actual observed outcomes adjusted by expected future loss due to non-optimal treatment. This approach prevents bias accumulation from stage to stage as compared to only using model based estimates. 

We can estimate the APP-pseudo-outcome recursively as 
\begin{equation} \label{e:pseudo-outcome}
\widetilde{PO}_{j} = U(F_j,S_j; \tilde{h}_j)+\sum_{t=j+1}^{T} \{U(F_t, S_t; \tilde{h}_t)+ \hat{\mu}_{t,g_t^{opt}}(\boldXbar_{t},\boldWbar_{t})- \hat{\mu}_{t, A_{t}}(\boldXbar_{t},\boldWbar_{t})\}    
\end{equation}

where $\hat{\mu}_{t,g_t^{opt}}(\boldXbar_{t},\boldWbar_{t})- \hat{\mu}_{t, A_{t}}(\boldXbar_{t},\boldWbar_{t})$ is the expected cumulative loss from stage $j$ onwards of not following the optimal regime during stage $t$. Both $\hat{\mu}_{t,g_t^{opt}}(\boldXbar_{t},\boldWbar_{t})$ and $ \hat{\mu}_{t, A_{t}}(\boldXbar_{t},\boldWbar_{t})$ can come from the same prediction model, which can take many forms, commonly parametric regression or random forests. Then, the proposed APP-AIPW estimator for $\hat{E}\{R_{j}^*(g_j)\}$ is $\mathbb{P}_n \left[\frac{I(A_{j}=g_j)}{\hat{\pi}_{j,A_j}(\boldXbar_{j},\boldWbar_{j})}\widetilde{PO}_{j} + \{1- \frac{I(A_{j}=g_j)}{\hat{\pi}_{j,A_j}(\boldXbar_{j},\boldWbar_{j})}\} \hat{\mu}_{j,g_j}(\boldXbar_{j},\boldWbar_{j}) \right]$, where $\widetilde{PO}_{j}$ takes the place of $R_{T}$ from stage $T$ and $\hat{\mu}_{j,g_j}(\boldXbar_{j},\boldWbar_{j})$ can be any model of $\widetilde{PO}_{j}$ that uses accumulated information up to stage $j$.

The propensity score $\pij(\boldXbar_{j}, \boldWbar_{j})$ for all stages $j$ can be estimated via multinomial logistic regression. In the above estimating equations, each $R_{j}$ requires an estimated $\tilde{h}_{j}$. Although for the sake of generality we suggest that $\pi$ and $\mu$ functions can be functions of both $\boldXbar_{j}$ and $\boldWbar_{j}$, we envision that $\boldWbar_{j}$ will mostly contribute through its effect $\tilde{h}_{j}$. We will discuss our proposed method to estimate $h_{j}$ in the next subsection.

\subsubsection{Implementation to obtain APP-weights}~\\

We assume going forward that information given by $\boldW_{j}$ will dominate information obtained from previous surveys (i.e. $\boldWbar_{j-1}$), and other covariate information ($\boldXbar_{j}$). Hence, following \cite{butler2018incorporating}, we assume here that $H_{j} \perp \boldXbar_{j} \vert \boldW_{j}$. This assumption can be weakened at the sake of more complicated models, which would be more burdensome to implement but follow an identical approach. Furthermore, we assume a latent traits model \citep{moustaki2000generalized} and that latent patient preferences are connected to items on the questionnaire through modified Rasch model \citep{rasch1960studies, rasch1961general}.

We assume the underlying generating form for a binary response is of $logit\{P(W_{jl}=1 \vert H_{j}=h_{j})\} = \alpha_{l0}+\alpha_{l1}h_{j}$ where $j$ is for stage and $l$ for the question number. If we wanted to relax the assumption that $H_{j}$ does not depend on $\boldXbar_{j}$ given $\boldWbar_{j}$, we can for example use the model $logit\{P(W_{jl}=1 \vert H_{j}=h_{j}, \boldX_{j} = \mathbf{x}_{j})\} = \alpha_{l0}+\alpha_{l1}h_{j} + \boldsymbol{\gamma}_l^T\mathbf{x}_{j}$, giving explicit dependence on covariate information. For cases where questions had more than two possible responses (e.g. three per question), assuming the first category is the reference (i.e. coefficients are $0$), we use the generating model $\log(\pi_{lb}(h_{j})/\pi_{la}(h_{j})) = \alpha_{lb0}+\alpha_{lb1}h_{j}$, and $\log(\pi_{lc}(h_{j})/\pi_{la}(h_{j})) = \alpha_{lc0}+\alpha_{lc1}h_{j}$. \\

\begin{algorithm}[H] \small \label{a: EMpreference}
\SetAlgoLined
\KwResult{Obtain $p(h_{j}\vert \boldw_{j})$ for patient $i$}
 Guess initial value of $h_j$ for all subjects to estimate an initial guess of $\boldsymbol{\alpha}_0, \boldsymbol{\alpha}_1$\;
 \While{not reached convergence}{
  Using MH, get an updated estimate of $p(h_j \vert \boldw_j)$\;
  Approximate likelihood integral using Gauss-Hermite quadrature with $k$ abscissae $h_t$ and weights $p(h_t)$;\\
  Solve likelihood equations using Newton-Raphson to get updated estimates of $\boldsymbol{\alpha}_0, \boldsymbol{\alpha}_1$\;
 }
 \caption{EM algorithm for estimating patient preference $\hat{h}_{j}$ }
\end{algorithm}

Algorithm \ref{a: EMpreference} outlines the algorithm for estimating patient preference $\hat{h}_{j}$. The APP-weights we propose are $\Phi(\hat{h}_{j})$. Essentially, the Expectation-Maximization algorithm \citep{moon1996expectation} iterates between estimates of $\boldsymbol{\alpha}$, the questionnaire coefficients, and $h_{j}$, individual patient preferences at stage $j$. In the process of derivation, we will use Gauss-Hermite quadrature to approximate the integral numerically, and estimate $P(h_{j} \vert \boldW_{j})\propto P(\boldW_{j} \vert h_{j})P(h_{j})$ through the Metropolis Hastings algorithm.

\subsubsection{APP-purity measure}~\\
The APP-purity measure that we propose is the following: \\$\mathcal{P}_j(\Omega, \omega)= max_{a_1, a_2 \in \mathcal{A}_j} \mathbb{P}_n \left[\summationj \muaipwhat(\boldXbar_{j},\boldWbar_{j})I\{g_{j,\omega,a_1, a_2}(\boldXbar_{j},\boldWbar_{j})=a_j \}I\{ \boldXbar_{j}\in \Omega\} \right]$. The node $\Omega$ here is the space that divides each individual in our dataset, and can be a factor of all observed data (in our case, $\Omega$ will depend on patient covariates $\boldXbar$). For a given partition $\omega$ and $\omega^c$ of node $\Omega$, $g_{j,\omega,a_1, a_2}$ denotes the decision rule that assigns treatment $a_1$ to subjects in $\omega$ and treatment $a_2$ to subjects in $\omega^c$ at stage $j (T \leq j \leq 1)$. Then, the two treatments $a_1$ and $a_2$ that yield the largest purity measure are selected to constitute the purity measure. Finally, the APP-RL algorithm uses this purity measure at each node to decide whether to split the tree. 

\section{Algorithmic Implementation} \label{s:algorithm}
As previously mentioned, $\mathcal{P}_j(\Omega, \omega)$ is the APP-purity measure of a potential split assigning treatment $a_1$ to patients in $\omega$, and $a_2$ to patients in $\omega^c$. Equivalently, $\mathcal{P}_j(\Omega)$ is the APP-purity measure when everyone in the node is assigned the single best treatment. The difference between $\mathcal{P}_j(\Omega, \omega)$ and $\mathcal{P}_j(\Omega)$ will provide primary guidance on if and how the node of the tree should split. 

To prevent overfitting, $\lambda$ is given to represent threshold for practical significance, and $n_0$ is given as minimal node size. The choice of $\lambda$ can be obtained through cross-validation or using domain knowledge, and $n_0$ could be selected a priori.

Under this set-up, we propose the following stopping rules:
\begin{enumerate}
    \item If node size is less than $2n_0$, the node will not be split
    \item If all possible splits of a node result in a child node with size smaller than $n_0$, the node will not be split
    \item If maximum purity improvement $\mathcal{P}_j(\Omega, \hat{w}^{opt})-\mathcal{P}_j(\Omega)$ is less than $\lambda$, where $\hat{w}^{opt} = arg\ max_{\omega} \left[ \mathcal{P}_j(\Omega,\omega) : min\{n\mathbb{P}_n I(\boldXbar_{j} \in \omega), n \mathbb{P}_n I(\boldXbar_{j} \in \omega^c) \}\geq n_0\right]$, the node will not be split. 
    \item If the current tree depth reaches the user-specified maximum depth, the tree growing process will stop
    \item Finally, if none of the stopping rules were triggered, split $\Omega$ into $\omega$ and $\omega^c$.
\end{enumerate}

This process is repeated at each node $\Omega$, until all of the potential nodes are terminated by a stopping rule. Note that $\mathcal{P}_j(\Omega)$ at each terminal node is also the expected counterfactual utility outcome (or expected APP-pseudo-outcome for stage $j<T$), which takes into account our chosen utility, observed toxicity and side effects, weighted by patient preference. The final tree will therefore use patient characteristics to assign each patient to a terminal node, which will determine their optimal stage specific decision that maximizes the preference weighted counterfactual utility outcome.  

\begin{algorithm}[H] \small \label{a: full alg}
\SetAlgoLined
\KwResult{$\mathbf{g^{opt}}=(g_1^{opt},\ldots,g_T^{opt})$}
 Initialize stage $j=T$\;
 \While{Stage $j \geq 1$}{
  Estimate $\hat{h}_{j}$ for each patient from $\boldW_{j}$ (all patients) using EM (See Alg. \ref{a: EMpreference})\;
  \eIf{$j=T$}{
   Combine $\hat{h}_{T}, F_{T}, S_{T}$ into $R_{T}$ \;
   Obtain $\hat{\pi}_{T,a_T}(\boldXbar_{T},\boldWbar_{T}), \hat{\mu}_{T,g_T}(\boldXbar_{T},\boldWbar_{T})$ and combine with $R_{T}$ to obtain $\hat{\mu}^{AIPW}_{T,a_T}(\boldXbar_{T},\boldWbar_{T})$ \;
    Set $m=1$ at root node $\Omega_{T,m}$ \;
   At node $\Omega_{T,m}$, evaluate the \textit{Stopping Rules}. If stop, assign the best treatment 
   $arg \max_{a_T\in \mathcal{A}_T} \mathbb{P}_n \left[ \hat{\mu}^{AIPW}_{T,a_T}(\boldXbar_{T},\boldWbar_{T})I\{ \boldXbar_{T}\in \Omega_{T,m}\} \right]$. Otherwise, split $\Omega_{T,m}$ into child nodes $\Omega_{T,2m}$ and $\Omega_{T,2m+1}$ by $\hat{\omega}^{opt}$ \;
   Set $m=m+1$ and repeat until all nodes are terminal $\xrightarrow {} \text{Obtain } g_T^{opt}$ \;
   }{
   Combine $\hat{h}_{j}, F_{j}, S_{j}$ into $R_{j}$ \;
   Estimate $\hat{\pi}_{j,a_j}(\boldXbar_{j},\boldWbar_{j})$. Using $g_{j+1}^{opt},\ldots,g_T^{opt}$ estimated previously, estimate $\hat{\mu}_{j,g_j}(\boldXbar_{j},\boldWbar_{j})$ and $\widetilde{PO}_{j}$. Combine together to obtain $\muaipwhat(\boldXbar_{j},\boldWbar_{j})$ \;
   Set $m=1$ at root node $\Omega_{j,m}$ \;
   At node $\Omega_{j,m}$, evaluate the \textit{Stopping Rules}. If stop, assign the best treatment 
   $arg \max_{a_j\in \mathcal{A}_j} \mathbb{P}_n \left[ \muaipwhat(\boldXbar_{j},\boldWbar_{j})I\{ \boldXbar_{j}\in \Omega_{j,m}\} \right]$. Otherwise, split $\Omega_{j,m}$ into child nodes $\Omega_{j,2m}$ and $\Omega_{j,2m+1}$ by $\hat{\omega}^{opt}$ \;
   Set $m=m+1$ and repeat until all nodes are terminal $\xrightarrow {} \text{Obtain } g_j^{opt}$ \;
   }
   Set $j=j-1$\;
 }
 \caption{APP-RL Algorithm Implementation}
\end{algorithm}

Algorithm \ref{a: full alg} provides a pseudo-code schematic for the layout of the entire algorithm. As we can see, the algorithm follows a backward induction strategy. At each stage, the algorithm updates patient preference estimates to those most recent and combines with observed outcomes of that stage into an APP-AIPW estimator of the expected counterfactual reward for each stage. The APP-AIPW estimator is then fed into the tree-based reinforcement learning algorithm to get the stage specific optimal rule. 

\section{Numerical Demonstration} \label{s:simulation}

We conduct simulation studies to investigate the performance of our proposed method.  We first consider a single-stage scenario with two treatments to facilitate comparison with Butler et al\cite{butler2018incorporating} in section \ref{scenario1}. Then, we also simulate a one-stage with three treatments in section \ref{scenario2} to evaluate how sensitive the performance of our proposed method depends on the number of treatment options, and a multi-stage with three treatments per stage in section \ref{scenario3} to assess the performance of the proposed method in multiple stage case. For all scenarios, we generate five independent baseline covariates $X_1,\ldots,X_5 \sim N(0,1)$. 

We simulate questionnaire responses as well as efficacy and side-effect outcomes for each individual. We then estimate both the patient preferences and optimal dynamic treatment from these responses and outcomes. For one stage, we obtain the percentage of subjects correctly classified to their true optimal treatment as $\%opt$. For multi-stage, $\%opt$ represents the percentage of subjects correctly classified to their true optimal regime (correct treatment for all stages).

\subsection{Scenario 1: $T=1$ and $K=2$} \label{scenario1}

In Scenario 1, we consider one-stage, two treatment options, and sample sizes of $300$, $500$, and $1000$. The observed treatment $A$ was generated from a $Bernoulli(\pi)$, where $\pi = \exp(0.5X_1+0.5X_4)/(1+\exp(0.5X_1+0.5X_4))$. 
The true underlying optimal rule is defined by:
\begin{equation*}
    g^{opt}(\mathbf{H})
    \begin{cases}
      1, & X_1 > -0.25, X_2 \leq 0.5 \\
      0, & \text{otherwise}
    \end{cases}
\end{equation*}

We generate the observed outcomes for efficacy and side effect ($F$ and $S$, respectively) as $F = 1+X_4+qf*(g^{opt}==A)+\epsilon_1$, and $S = 1+X_5+qs*(g^{opt}==A)+\epsilon_2$. To ensure that the distributions of $F$ and $S$ are similar, a $Bernoulli(0.5)$ random variable was used to determine whether $qs$ or $qf$ would be simulated first and set to $3X_3$. The latter was set to $(-\Phi(h)*qf/(1-\Phi(h)))+(1/(1-\Phi(h)))$ or $((\Phi(h)-1)*qs/\Phi(h)+(1/\Phi(h))$ for $qs$ and $qf$, respectively, where $\Phi(\cdot)$ is the cumulative density distribution of a standard normal variable. This complicated form for $qf,qs$ simply ensures that the overall reward $R=\Phi(h)F+\{1-\Phi(h)\}S$ is greater by $1$ when the correct treatment is chosen. The concordance between optimal regime for $F$ and $S$ was approximately $25-30\%$, indicating that for over $75\%$ of cases, there is a trade-off between side-effect and efficacy. 

We looked at the performance of our method under two different patient preference distributions. Patient preferences were generated from either the $Uniform(-1,1)$ or $0.5Normal(X_2, SD=0.2)$ distribution, where preference depended on an underlying covariate. Each patient answered a questionnaire of ten questions, and each question allowed a binary $\{0,1\}$ response. For each question $l$ we assume a latent traits model, where $W_{jl} \sim Bernoulli\{expit(\alpha_{l0}+\alpha_{l1}H_j)\}$. The coefficients for each of the questions for Scenario $1$ is in Table \ref{t:scenario1_combined} Panel A. 

In calculating the APP-purity measure, we specified the $\pi-$model with covariates $X_1$ and $X_4$ used to assign the treatment. Similarly, we specified the $\mu-$model in the augmentation term using a linear regression framework with $R=\Phi(\hat{h})F+\{1-\Phi(\hat{h})\}S$ as the outcome, and treatment indicator $A$, patient characteristics $X_1,\ldots,X_5$ and their interaction terms with $A$ as covariates. 
Table \ref{t:scenario1_combined} Panel B shows results of simulation for the methods in Scenario $1$. We compared our method against that proposed in Butlerl et al \cite{butler2018incorporating}. As suggested in their paper, we fit linear working models with all covariates, questionnaire responses, and their interaction terms with treatment. For $n=300$ and when $h$ followed an uniform distribution, APP-RL method was able to select $87.1\%$ of optimal treatment, which increased to $96.6\%$ when $n=1000$. In contrast, the Q-learning method lags behind in accuracy by over $15\%$, at $70.6\%$ and $77.3\%$ respectively. However, both methods saw reduction in standard errors as sample size increased. The patterns are similar when patient preference was generated with a normal distribution. 

\begin{table}[ht]
\centering
\caption{Scenario 1 analysis: Latent traits model coefficients and simulation results} 
\label{t:scenario1_combined}
\vspace{0.5em}
\begin{threeparttable}
% ------------------- PANEL A (LEFT) -------------------
\begin{minipage}[t]{0.35\textwidth}
  \parbox{0.85\textwidth}{\centering Panel A: Questionnaire coefficients for the latent traits model}\\[0.8em]
  \centering
  \begin{tabularx}{\linewidth}{|X|X|X|}
    \toprule
    $l$ & $\alpha_0$ & $\alpha_1$ \\ 
    \midrule
    1  & -0.25 & 0.92 \\ 
    2  & -0.84 & 0.64 \\ 
    3  & -1.64 & 2.35 \\ 
    4  & -0.78 & 0.82 \\ 
    5  & -0.89 & 1.10 \\ 
    6  &  0.42 & 0.16 \\ 
    7  &  1.27 & 2.96 \\ 
    8  & -0.61 & 0.56 \\ 
    9  &  0.09 & 0.30 \\ 
    10 & -0.40 & 1.35 \\ 
    \bottomrule
  \end{tabularx}
\end{minipage}%
\hfill
% ------------------- PANEL B (RIGHT) -------------------
\begin{minipage}[t]{0.6\textwidth}
  \centering
    Panel B: \% optimal chosen for 1-stage, 2 treatment, binary responses (200 iterations). Setting 1 refers to $h \sim U(-1,1)$, and Setting 2 refers to $h \sim 0.5 N(X_2, \text{SD}=0.2)$\\[0.5em]
    
  \begin{tabular}{@{}llccc@{}}
    \toprule
    & & $n=300$ & $n=500$ & $n=1000$ \\ 
    \cmidrule(lr){3-5}
    & & \% opt (SD) & \% opt (SD) & \% opt (SD) \\ 
    \midrule
    Setting 1 & APP-RL     & 87.06 (9.70)  & 92.65 (6.88) & 96.56 (4.76) \\
              & Q-learning & 70.64 (4.12)  & 73.75 (3.63) & 77.26 (2.20) \\ 
    \midrule
    Setting 2 & APP-RL     & 87.86 (10.45) & 90.65 (8.57) & 96.44 (5.35) \\
              & Q-learning & 71.12 (5.62)  & 73.71 (4.49) & 75.82 (3.74) \\ 
    \bottomrule
  \end{tabular}
  % ------------------- TABLE NOTES -------------------
\begin{tablenotes}
  \small
  \item APP-RL: Augmented Patient Preference incorporated Reinforcement Learning. Q-learning refers to the method by Butler et al \cite{butler2018incorporating}.
\end{tablenotes}
\end{minipage}
\vspace{0.5em}
\end{threeparttable}
\end{table}

\subsection{Scenario 2: $T=1$ and $K=3$} \label{scenario2}

In this scenario, we consider one-stage with three treatment options. We again investigate sample sizes of $300$, $500$, and $1000$. In this case, treatment $A$ could take values in $\{0,1,2\}$ generated from $Multinomial(\boldsymbol{\pi})$, where $\boldsymbol{\pi} = [\pi_0, \pi_1, \pi_2] = [1/(1+\exp(0.5X_4+0.5X_1)+\exp(0.5X_5-0.5X_1)),\exp(0.5X_4+0.5X_1)/ (1+\exp(0.5X_4+0.5X_1)+\exp(0.5X_5-0.5X_1)), 1-\pi_0-\pi_1]$.

The true underlying optimal rule is defined by: 
\begin{equation*}
    g^{opt}(\mathbf{H})
    \begin{cases}
      0, & X_1 \leq 0, X_2 \leq 0.5 \\
      2, & X_1 > 0, X_3 \leq 0.5 \\
      1, & \text{otherwise}
    \end{cases}
\end{equation*}

$F$ and $S$ were generated in the same way as in Scenario $1$, but the first of $qf$ or $qs$ was set to $2X_3-(1.25X_2)^2$. The second variable ($qs/qf$) was generated similarly to that of stage $1$ to ensure an advantage of $1.5$ on the reward if the optimal treatment was chosen.  

Patient preferences were likewise generated from either an uniform $Uniform(-1,1)$ or a $0.5Normal(X_2, SD=0.2)$ distribution, and each patient answered ten questions with three categorical responses in a survey. For each question $l$, assuming the first category is the reference (i.e. coefficients are $0$), we used the generating model $\log(\pi_{lb}(h)/\pi_{la}(h)) = \alpha_{lb0}+\alpha_{lb1}h$, and $\log(\pi_{lc}(h)/\pi_{la}(h)) = \alpha_{lc0}+\alpha_{lc1}h$. Hence, the vector of probabilities we feed into a multinomial random generator to generate $\boldW_{l}$ is: \\
$\boldsymbol{\pi}_{l}(h)=[\pi_{la}(h),\pi_{lb}(h),\pi_{lc}(h)]$, where \\
$\pi_{la}(h) = 1/\{1+\exp(\alpha_{lb0}+\alpha_{lb1}h)+\exp(\alpha_{lc0}+\alpha_{lc1}h)\}$, \\
$\pi_{lb}(h) = \{\exp(\alpha_{lb0}+\alpha_{lb1}h)\}/\{1+\exp(\alpha_{lb0}+\alpha_{lb1}h)+\exp(\alpha_{lc0}+\alpha_{lc1}h)\}$, and \\
$\pi_{lc}(h) = \{\exp(\alpha_{lc0}+\alpha_{lc1}h)\}/\{1+\exp(\alpha_{lb0}+\alpha_{lb1}h)+\exp(\alpha_{lc0}+\alpha_{lc1}h)\}$. For each person and each question, we generate $\mathbf{W}_{l} \sim Multinomial(\pi_{la}(h),\pi_{lb}(h),\pi_{lc}(h))$.

The exact coefficients for each of the questions for Scenario $2$ is shown below in Table \ref{t:scenario2_combined} Panel A.

We estimated the working models in an equivalent way to Scenario $1$. Table \ref{t:scenario2_combined} Panel B showcases results from our APP-RL method. Because the Q-learning based method of Butler et al. \cite{butler2018incorporating} is unable to handle more than two treatments and more than two responses per question, it was not possible to directly compare with our method. In general, we see that the algorithm does better at selecting the optimal regime given a smaller range of patient preference (i.e. the uniform distribution), but the difference is small. We can also see that both accuracy and efficiency increases as sample size increases, with over $83\%$ selected as optimal for $n=300$, and over $95\%$ for $n=1000$. 

\begin{table}[htbp]
\centering
\caption{Scenario 2 analysis: Latent traits model coefficients and simulation results} 
\label{t:scenario2_combined}
\vspace{0.5em}
\begin{threeparttable}

% ------------------- PANEL A (LEFT) -------------------
\begin{minipage}[t]{0.42\textwidth}
  \centering
  Panel A: Questionnaire coefficients for the latent traits model\\[0.5em]
  \begin{tabularx}{\linewidth}{|X|X|X|X|X|}
    \toprule
    $l$ & $\alpha_{b0}$ & $\alpha_{b1}$ & $\alpha_{c0}$ & $\alpha_{c1}$ \\ 
    \midrule
    1  &  0.19 &  1.61 &  0.15 & 1.76 \\ 
    2  &  0.01 & -0.51 & -0.47 & 3.12 \\ 
    3  & -0.77 &  2.87 &  0.38 & 2.49 \\ 
    4  &  0.68 &  1.08 &  0.83 & 1.74 \\ 
    5  & -0.77 & -0.30 &  0.74 & 0.84 \\ 
    6  & -0.48 &  0.24 &  0.49 & 0.72 \\ 
    7  &  0.23 &  2.26 &  0.83 & 1.48 \\ 
    8  &  0.91 &  0.33 &  0.89 & 2.01 \\ 
    9  &  0.72 & -0.12 & -0.56 & 0.70 \\ 
    10 &  0.88 &  0.84 & -0.30 & 0.66 \\ 
    \bottomrule
  \end{tabularx}
\end{minipage}%
\hfill
% ------------------- PANEL B (RIGHT) -------------------
\begin{minipage}[t]{0.54\textwidth}
  \centering
  Panel B: \% optimal chosen for 1-stage, 3 treatment, 3-responses (200 iterations). Setting 1 refers to $h \sim U(-1,1)$, and Setting 2 refers to $h \sim 0.5 N(X_2, \text{SD}=0.2)$\\[0.5em]
  \begin{tabular}{@{}lccc@{}}
    \toprule
    & $n=300$ & $n=500$ & $n=1000$ \\ 
    \cmidrule(lr){2-4}
    & \% opt (SD) & \% opt (SD) & \% opt (SD) \\ 
    \midrule
    Setting 1 & 85.26 (13.72) & 92.74 (7.81)  & 97.48 (4.47) \\ 
    Setting 2 & 82.72 (15.79) & 90.04 (11.39) & 95.18 (7.94) \\ 
    \bottomrule
  \end{tabular}
\end{minipage}

\vspace{0.8em}
\end{threeparttable}
\end{table}

\subsection{Scenario 3: $T=2$ and $K_1=K_2=3$} \label{scenario3}
In this scenario, we consider a two-stage set-up, with three treatment options at each stage and 3-category response per question on the survey. We simulated sample sizes of $1000$ and $2000$. The outcome to be maximized is the sum of expected rewards of each stage. As in scenario 2, treatment $A_1$ could take values in $\{0,1,2\}$ generated from $Multinomial(\boldsymbol{\pi})$, where $\boldsymbol{\pi} = [\pi_0, \pi_1, \pi_2] = [1/(1+\exp(0.5X_4+0.5X_1)+\exp(0.5X_5-0.5X_1)),(\exp(0.5X_4+0.5X_1)/ (1+\exp(0.5X_4+0.5X_1)+\exp(0.5X_5-0.5X_1)), 1-\pi_0-\pi_1]$.

For this scenario, we looked at two forms of true underlying rules, tree-type and non-tree type. The stage $1$ tree-type optimal regime is defined by $g_1^{opt} = I(X_1>-0.65)\{I(X_2>-0.75)+I(X_2>0.1)\}$, while the non-tree type is defined by $g_1^{opt} = I(X_1>-0.3)\{1+I(X_1+X_2>0.3)\}$. We also investigated performance under both equal (where selection of the optimal treatment guarantees a uniform reward advantage over the other two treatment options) and varying penalty setting (where selection of the optimal treatment has differing reward advantages as compared to the other treatments selected). For the equal penalty setting, $F_1 = 1+X_4+0.7X_1+qf\cdot I(A_1 \neq g_1^{opt})+\epsilon_1$ and $S_1 = 1+X_5+0.7X_1+qs\cdot I(A_1 \neq g_1^{opt})+\epsilon_2$. For the varying penalty scenario, $F_1 = 1+X_4+1.3X_1+qf\cdot \left|A_1- g_1^{opt}\right|+\epsilon_1$ and $S_1 = 1+X_5+1.3X_1+qs\cdot \left|A_1- g_1^{opt} \right|+\epsilon_2$. As before, the first of $qs$ and $qf$ to be simulated was set to $2X_3+(1+2X_5)^2$, and the latter was set to $-\phi(h_{1})*qf/(1-\phi(h_{1})-(2.25/(1-\phi(h_{1}))$ and 
$(\phi(h_{1})-1)*qs/\phi(h_{1})-2.25/\phi(h_{1})$, for $qs$ and $qf$ respectively, where $\phi(\cdot)$ is the cumulative density distribution for a $N(0,SD=3)$ random variable. 

Stage $2$ parameters followed the same pattern as those from stage $1$. Treatment $A_2$ could take values in $\{0,1,2\}$ generated from $Multinomial(\boldsymbol{\gamma})$, where $\boldsymbol{\gamma} = [\gamma_0, \gamma_1, \gamma_2] = [1/(1+\exp(0.2U_1-0.5)+\exp(0.5X_2)),(\exp(0.2U_1-0.5)/ (1+\exp(0.2U_1-0.5)+\exp(0.5X_2)), 1-\gamma_0-\gamma_1]$, where $U_1$ is the utility of stage 1.  The stage $2$ tree-type optimal rule is defined by $g_2^{opt} = I(X_2>-0.5)\{I(U_1>-7)+I(U_1>0)\}$, while the non-tree type is defined by $g_1^{opt} = I(X_2>0.15)\{1+I(X_2+U_1>1.25)\}$. For the equal penalty setting, $F_2 = 1+1.5X_3+qf\cdot 0.7I(A_2 \neq g_2^{opt})+\epsilon_1$ and $S_2 = 1+1.5X_3+qs\cdot0.7 I(A_2 \neq g_2^{opt})+\epsilon_2$. For the varying penalty setting, $F_2 = 1+1.5X_3+qf\cdot \left|A_2- g_2^{opt}\right|+\epsilon_1$ and $S_2 = 1+1.5X_3+qs\cdot \left|A_2- g_2^{opt} \right|+\epsilon_2$. As before, the first of $qs$ and $qf$ to be simulated was set to $1.25X_4+(1.7X_1)^2$, and the latter was set to $-\phi(h_{2})*qf/(1-\phi(h_{2}))-(1.5/(1-\phi(h_{2}))$ and 
$(\phi(h_{2})-1)*qs/(\phi(h_{2}))-(1.5/(1-\phi(h_{2}))$, for $qs$ and $qf$ respectively. 

For both stages, the concordance between $F$ and $S$ was approximately $17-20\%$, indicating for vast majority of cases, the best treatment for maximizing efficiency was not the same as the best treatment for minimizing toxicity. 

Patient preferences at stage $1$ was generated from a $Normal(0,SD=0.5)$ distribution for all subjects. Stage $2$ preferences were generated from $0.3Normal(U_1, SD=0.4)$ distribution, thereby assuming that stage $2$ preferences are influenced by stage $1$ outcomes and satisfaction levels. We assume the same latent model for questionnaire parameters as in scenario 2. The first stage coefficients are identical to those in scenario 2, and the stage 2 questionnaire coefficients are shown in Table \ref{t:scenario3_combined} Panel A. 

\begin{table}[htbp]
\centering
\caption{Scenario 3 analysis: Latent traits model coefficients and simulation results} 
\label{t:scenario3_combined}
\vspace{0.5em}
\begin{threeparttable}
% ------------------- PANEL A (LEFT) -------------------
\begin{minipage}[t]{0.4\textwidth}
  \centering
  Panel A: Stage 2 questionnaire coefficients for the latent traits model\\[0.5em]
  \begin{tabularx}{\linewidth}{|X|X|X|X|X|}
    \toprule
$l$ & $\alpha_{b0}$ & $\alpha_{b1}$ & $\alpha_{c0}$ & $\alpha_{c1}$ \\ 
  \midrule
1 & -0.09 & 1.10 & -0.00 & -0.28 \\ 
  2 & -0.52 & 0.85 & -1.64 & -0.62 \\ 
  3 & -1.69 & 1.76 & -0.43 & -0.52 \\ 
  4 & 1.86 & 1.74 & 0.45 & 1.20 \\ 
  5 & -2.08 & 1.36 & 0.07 & -2.95 \\ 
  6 & 2.18 & 1.32 & 0.52 & 1.01 \\ 
  7 & 0.73 & 1.38 & -0.84 & -3.07 \\ 
  8 & -0.60 & 2.14 & -1.93 & -1.30 \\ 
  9 & 0.69 & 0.86 & 0.67 & -1.62 \\ 
  10 & -0.29 & 1.08 & -1.81 & -2.20 \\ 
   \bottomrule
  \end{tabularx}
\end{minipage}%
\hfill
% ------------------- PANEL B (RIGHT) -------------------
\begin{minipage}[t]{0.54\textwidth}
  \centering
  Panel B: \% optimal regimes chosen for 2-stage, with 3 treatments options in each stage, with 200 iterations.
\begin{tabular}{lll}
\toprule
Simulation Structure           & n=1000        & n=2000        \\ \midrule
                           & \% opt (SD)   & \% opt (SD)   \\
Tree and equal penalty & 91.6 (2.89)   & 93.5 (1.53)   \\
Tree and varying penalty & 90.2 (5.33)   & 92.7 (3.20)   \\
Nontree and equal penalty & 81.6 (3.88)   & 82.3 (3.60)   \\
Nontree and varying penalty & 83.3 (4.40)   & 84.5 (3.96)   \\ \bottomrule
\end{tabular}
\begin{tablenotes}
      \small
      \centering
      \item Underlying true simulated structure was generated as either a tree or non-tree structure. Equal penalty guarantees a uniform reward advantage for the optimal treatment over other treatments, whereas varying penalty structure does not. 
\end{tablenotes}
\end{minipage}
\end{threeparttable}
\end{table}

Model specification for stage 2 involves specifying the $\pi_2$-model using $\hat{U}_1$ and $X_2$. The $\mu_2-$ model in the augmentation term also uses $\hat{U}_2$ as the outcome, and $A_1,A_2$ and patient characteristics $X_1,\ldots,X_5, \hat{U}_1$, and the interaction terms between $A_2$ and all the rest of the terms as covariates. This set up allows us to obtain the optimal decision tree for stage 2. The $\pi_1$-model uses $X_1,X_4$, and $X_5$ as covariates. In calculating the pseudo-outcome $\widetilde{PO}_1=\hat{U}_1+\hat{U}_2+\hat{\mu}_{2,g_2^{opt}}(\boldXbar_2)-\hat{\mu}_{2,A_2}(\boldXbar_2)$, we obtain both $\hat{\mu}_{2,g_2^{opt}}(\boldXbar_2)$ and $\hat{\mu}_{2,A_2}(\boldXbar_2)$ through predictions obtained through a random forest model with $A_1,A_2,X_1,\ldots,X_5,\hat{U}_1$ as covariates, and $\hat{U}_2$ as the outcome. Finally, the $\mu_1$-model in the first stage augmentation term used $A_1,X_1,\ldots,X_5$ and the interaction terms between $A_1$ and the patient characteristics as covariates. 

\begin{figure}
    \centering
    \includegraphics[width=1\linewidth]{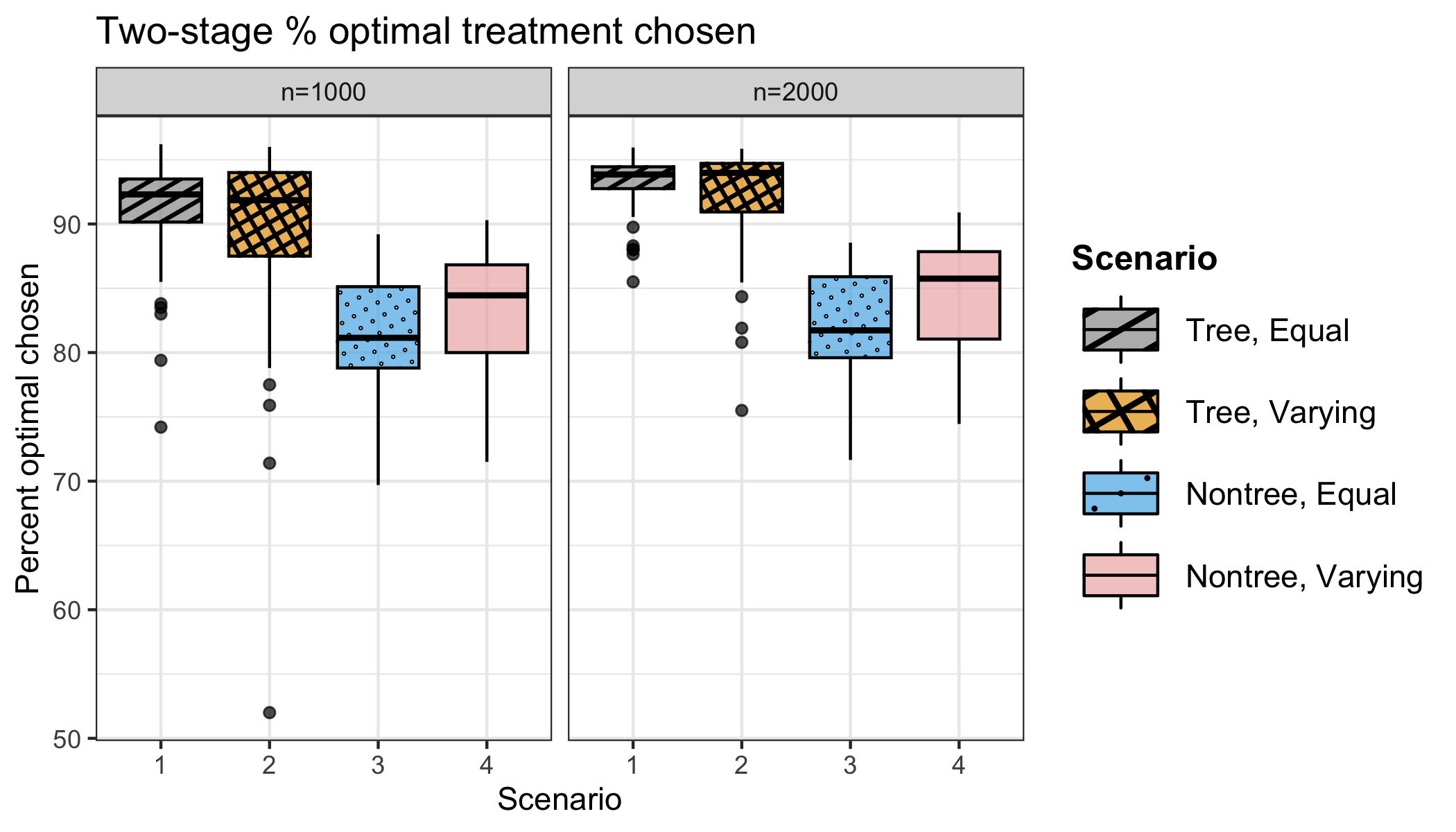}
    \caption{Results for 2-stage, 200 iterations. We looked at the case where the true underlying reward structure follows a tree and non-tree structure, as well as the equal and varying penalty settings for the reward.} \label{f:scenario3}
\end{figure}

Table \ref{t:scenario3_combined} Panel B and Figure \ref{f:scenario3} show simulation results for each of the settings in this scenario. In all settings, our method was able to select the correct regimen (the correct treatment for both Stage 1 and for Stage 2) over $80\%$ of the time, much higher than that could have been gotten by random chance (approximately $(1/3)^2 = 1/9$). Furthermore, we can see that when the underlying distribution is tree-based, the method does $\sim 10\%$ better than when the underlying optimal regime is not tree-based. Additionally, we can see the method improves with increased sample size. 

\begin{figure}
    \centering
    \includegraphics[width=1\linewidth]{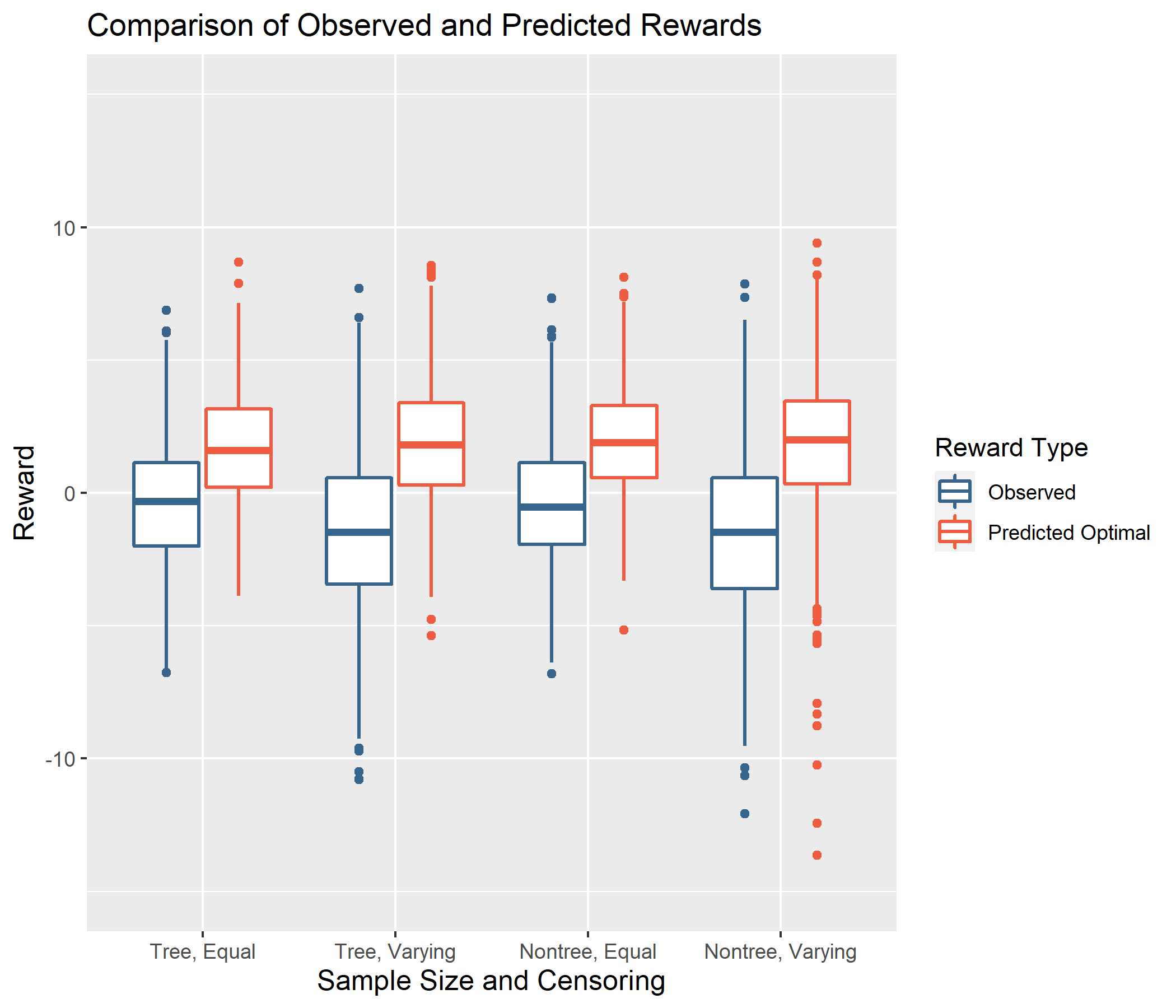}
    \caption{Comparison of observed cumulative reward vs predicted cumulative reward if following APP-RL's treatment predictions for the case of $n=1000$. Both rewards are calculated using simulated true preferences. }
    \label{apprl:rewards_plot}
\end{figure}

Figure \ref{apprl:rewards_plot} shows the observed rewards as compared with predicted rewards, where a patient hypothetically follows the predicted optimal treatments obtained through our algorithm. Both rewards are calculated using true patient preferences but predicted optimal treatment assignments were obtained using estimated patient preferences. In aggregate across the four scenarios, the median difference between predicted optimal reward and observed true reward is $2.722$, and over $85\%$ of patients on average derived benefit from following the predicted optimal treatment as compared to their observed treatment. 

\section{Discussion}\label{s:discussion}
In this report, we propose a method that estimates an optimal dynamic treatment regime that maximizes a patient preferred utility function using a tree-based reinforcement learning approach. The vast majority of dynamic treatment regime methods in the literature optimize a single outcome. The main published method that the authors are aware of with an endpoint that incorporates patient preference is work by Butler et al\cite{butler2018incorporating}. However, the authors wish to highlight a few key differences between the our proposed method and theirs. First, the method in Butler et al.\cite{butler2018incorporating} is for single stage scenarios and is not designed to handle multiple stage decision making. Secondly, their method is designed to select between two potential treatments, which is clinically limiting. Thirdly, their method estimates patient preference from a survey of questions with binary choices, which we have extended to surveys with categorical choices. In summary, our method brings in patient preference incorporation into the chronic disease, multi-stage, multiple treatment option setting. Furthermore, incorporation of APP-AIPW into the purity measure endows our model with the doubly robust feature, providing a safety net against model misspecification. 

As shown in Table \ref{t:scenario1_combined}, our model has strong performance across the board for both one-stage, binary treatment scenarios and outperforms traditional Q-learning approaches when the underlying functional form for benefit follows a tree-based structure. Similarly, we show promising results in Table \ref{t:scenario2_combined} for the one stage, three treatment options, where we obtain over $80\%$ for all settings. In both of these tables, we can see that the prediction accuracy of APP-RL increases with sample size and its associated variance decreases with increasing sample size. For the two-stage scenario, Figure \ref{f:scenario3} shows that our method does best when the underlying distribution is tree-type, which is unsurprising given the algorithmic architecture. However, it still does respectively well even when the underlying distribution is not tree-type, indicating that our method is generally robust and applicable to more than one setting. With tree-type distribution we see significant reduction in variation with increasing sample size, something we see less of with the non-tree distribution. In general, equal penalty seems to be associated with smaller variability, but the type of penalty seems to slightly influence the prediction accuracy of our method. In general, no matter the type of underlying distribution or type of penalty, the expected increase in rewards show similar patterns across the board, with the vast majority of cases gaining improved reward by following APP-RL's recommended treatment sequence. 

There are a number of potential improvements and extensions that we could explore in future studies. Generalizing our method and researching potential utility functions to accommodate more than two competing outcomes would be one improvement. Even more ambitious would be to move this augmentation patient preference framework into multi-objective optimization, where we could directly optimize in $n$-dimensional space. Instead of obtaining an unique solution, the goal would be to produce a set of non-dominated (where no one solution is better than the others in all ways) solutions. More precise and efficient ways of estimating patient preference would be of value, as the reliance on questionnaire and subsequent sampling and numerical methods are both labor and time intensive. Finally, the incorporation of continuous stages (i.e. mobile health interventions) which are heavily influenced by personal decisions would be of interest in this work, as it would bring together instantaneous preference estimation and decision making for more general and timely clinical scenarios.

\section{Declaration of conflicting interests}
The author(s) declared no potential conflicts of interest with respect to the research, authorship, and/or publication of this article.

\section{Funding}
The author(s) received no financial support for the research, authorship, and/or publication of this article.

\bibliographystyle{SageV}
\bibliography{ref}

\begin{thebibliography}{10}
\providecommand{\url}[1]{\texttt{#1}}
\providecommand{\urlprefix}{URL }
\expandafter\ifx\csname urlstyle\endcsname\relax
  \providecommand{\doi}[1]{DOI:\discretionary{}{}{}#1}\else
  \providecommand{\doi}{DOI:\discretionary{}{}{}\begingroup \urlstyle{rm}\Url}\fi
\providecommand{\eprint}[2][]{\url{#2}}

\bibitem{hamburg2010path}
Hamburg MA and Collins FS.
\newblock The path to personalized medicine.
\newblock \emph{New England Journal of Medicine} 2010; 363(4): 301--304.

\bibitem{national2011toward}
Council NR et~al.
\newblock \emph{Toward precision medicine: building a knowledge network for biomedical research and a new taxonomy of disease}.
\newblock National Academies Press, 2011.

\bibitem{chakraborty2013statistical}
Chakraborty B and Moodie E.
\newblock \emph{Statistical methods for dynamic treatment regimes}.
\newblock Springer, 2013.

\bibitem{chakraborty2014dynamic}
Chakraborty B and Murphy SA.
\newblock Dynamic treatment regimes.
\newblock \emph{Annual review of statistics and its application} 2014; 1: 447--464.

\bibitem{murphy2003optimal}
Murphy SA.
\newblock Optimal dynamic treatment regimes.
\newblock \emph{Journal of the Royal Statistical Society: Series B (Statistical Methodology)} 2003; 65(2): 331--355.

\bibitem{sutton2018reinforcement}
Sutton RS and Barto AG.
\newblock \emph{Reinforcement learning: An introduction}.
\newblock MIT press, 2018.

\bibitem{moodie2012q}
Moodie EE, Chakraborty B and Kramer MS.
\newblock Q-learning for estimating optimal dynamic treatment rules from observational data.
\newblock \emph{Canadian Journal of Statistics} 2012; 40(4): 629--645.

\bibitem{schulte2014q}
Schulte PJ, Tsiatis AA, Laber EB et~al.
\newblock Q-and a-learning methods for estimating optimal dynamic treatment regimes.
\newblock \emph{Statistical science: a review journal of the Institute of Mathematical Statistics} 2014; 29(4): 640.

\bibitem{robins2000marginal}
Robins JM.
\newblock Marginal structural models versus structural nested models as tools for causal inference.
\newblock In \emph{Statistical models in epidemiology, the environment, and clinical trials}. Springer, 2000.
\newblock pp. 95--133.

\bibitem{robins2004optimal}
Robins JM.
\newblock Optimal structural nested models for optimal sequential decisions.
\newblock In \emph{Proceedings of the second seattle Symposium in Biostatistics}. Springer, pp. 189--326.

\bibitem{thall2007bayesian}
Thall PF, Wooten LH, Logothetis CJ et~al.
\newblock Bayesian and frequentist two-stage treatment strategies based on sequential failure times subject to interval censoring.
\newblock \emph{Statistics in medicine} 2007; 26(26): 4687--4702.

\bibitem{laber2015tree}
Laber E and Zhao Y.
\newblock Tree-based methods for individualized treatment regimes.
\newblock \emph{Biometrika} 2015; 102(3): 501--514.

\bibitem{tao2017adaptive}
Tao Y and Wang L.
\newblock Adaptive contrast weighted learning for multi-stage multi-treatment decision-making.
\newblock \emph{Biometrics} 2017; 73(1): 145--155.

\bibitem{tao2018tree}
Tao Y, Wang L and Almirall D.
\newblock Tree-based reinforcement learning for estimating optimal dynamic treatment regimes.
\newblock \emph{The annals of applied statistics} 2018; 12(3): 1914.

\bibitem{zhang2018interpretable}
Zhang Y, Laber EB, Davidian M et~al.
\newblock Interpretable dynamic treatment regimes.
\newblock \emph{Journal of the American Statistical Association} 2018; 113(524): 1541--1549.

\bibitem{ortinski2004cognitive}
Ortinski P and Meador KJ.
\newblock Cognitive side effects of antiepileptic drugs.
\newblock \emph{Epilepsy \& Behavior} 2004; 5: 60--65.

\bibitem{krumholz2009driving}
Krumholz A.
\newblock Driving issues in epilepsy: past, present, and future.
\newblock \emph{Epilepsy Currents} 2009; 9(2): 31--35.

\bibitem{barry2012shared}
Barry MJ and Edgman-Levitan S.
\newblock Shared decision making—the pinnacle of patient-centered care.
\newblock \emph{New England Journal of Medicine} 2012; 366(9): 780--781.

\bibitem{basu2007value}
Basu A and Meltzer D.
\newblock Value of information on preference heterogeneity and individualized care.
\newblock \emph{Medical Decision Making} 2007; 27(2): 112--127.

\bibitem{kashaf2015does}
Kashaf MS and McGill E.
\newblock Does shared decision making in cancer treatment improve quality of life? a systematic literature review.
\newblock \emph{Medical decision making} 2015; 35(8): 1037--1048.

\bibitem{shay2015evidence}
Shay LA and Lafata JE.
\newblock Where is the evidence? a systematic review of shared decision making and patient outcomes.
\newblock \emph{Medical Decision Making} 2015; 35(1): 114--131.

\bibitem{oshima2013shared}
Oshima~Lee E and Emanuel EJ.
\newblock Shared decision making to improve care and reduce costs.
\newblock \emph{New England Journal of Medicine} 2013; 368(1): 6--8.

\bibitem{embretson2013item}
Embretson SE and Reise SP.
\newblock \emph{Item response theory}.
\newblock Psychology Press, 2013.

\bibitem{butler2018incorporating}
Butler EL, Laber EB, Davis SM et~al.
\newblock Incorporating patient preferences into estimation of optimal individualized treatment rules.
\newblock \emph{Biometrics} 2018; 74(1): 18--26.

\bibitem{moustaki2000generalized}
Moustaki I and Knott M.
\newblock Generalized latent trait models.
\newblock \emph{Psychometrika} 2000; 65(3): 391--411.

\bibitem{bartholomew2011latent}
Bartholomew DJ, Knott M and Moustaki I.
\newblock \emph{Latent variable models and factor analysis: A unified approach}, volume 904.
\newblock John Wiley \& Sons, 2011.

\bibitem{marler2004survey}
Marler RT and Arora JS.
\newblock Survey of multi-objective optimization methods for engineering.
\newblock \emph{Structural and multidisciplinary optimization} 2004; 26(6): 369--395.

\bibitem{lizotte2012linear}
Lizotte DJ, Bowling M and Murphy SA.
\newblock Linear fitted-q iteration with multiple reward functions.
\newblock \emph{Journal of Machine Learning Research} 2012; 13(Nov): 3253--3295.

\bibitem{murphy2001marginal}
Murphy SA, van~der Laan MJ, Robins JM et~al.
\newblock Marginal mean models for dynamic regimes.
\newblock \emph{Journal of the American Statistical Association} 2001; 96(456): 1410--1423.

\bibitem{orellana2010dynamic}
Orellana L, Rotnitzky A and Robins JM.
\newblock Dynamic regime marginal structural mean models for estimation of optimal dynamic treatment regimes, part i: main content.
\newblock \emph{The international journal of biostatistics} 2010; 6(2).

\bibitem{robins2009estimation}
Robins JM and Hern{\'a}n MA.
\newblock Estimation of the causal effects of time-varying exposures.
\newblock \emph{Longitudinal data analysis} 2009; 553: 599.

\bibitem{breiman2017classification}
Breiman L.
\newblock \emph{Classification and regression trees}.
\newblock Routledge, 2017.

\bibitem{rotnitzky1998semiparametric}
Rotnitzky A, Robins JM and Scharfstein DO.
\newblock Semiparametric regression for repeated outcomes with nonignorable nonresponse.
\newblock \emph{Journal of the american statistical association} 1998; 93(444): 1321--1339.

\bibitem{huang2015optimization}
Huang X, Choi S, Wang L et~al.
\newblock Optimization of multi-stage dynamic treatment regimes utilizing accumulated data.
\newblock \emph{Statistics in medicine} 2015; 34(26): 3424--3443.

\bibitem{rasch1960studies}
Rasch G.
\newblock \emph{Probabilistic models for some intelligence and attainment tests}, \emph{Studies in mathematical psychology}, volume~1.
\newblock Copenhagen: Nielsen \& Lydiche, 1960.

\bibitem{rasch1961general}
Rasch G.
\newblock On general laws and the meaning of measurement in psychology.
\newblock In \emph{Proceedings of the fourth Berkeley symposium on mathematical statistics and probability}, volume~4. pp. 321--333.

\bibitem{moon1996expectation}
Moon TK.
\newblock The expectation-maximization algorithm.
\newblock \emph{IEEE Signal processing magazine} 1996; 13(6): 47--60.

\end{thebibliography}

\end{document}